\documentstyle[aps,preprint,floats,psfig]{revtex}
\begin{document}
\newcommand {\be}{\begin{equation}}
\newcommand {\ee}{\end{equation}}
\newcommand {\bea}{\begin{eqnarray}}
\newcommand {\eea}{\end{eqnarray}}
\newcommand {\nn}{\nonumber}

\draft
%
%
%
%

\title{
Magnetic Field Induced Ordering in Quasi-One-Dimensional Quantum Magnets
}

\author{Stefan Wessel and Stephan Haas}
\address{Department of Physics and Astronomy, University of Southern
California, Los Angeles, CA 90089-0484}

\date{\today}
\maketitle

\begin{abstract}
Three-dimensional magnetic ordering
transitions are studied theoretically in strongly anisotropic 
quantum magnets.
An external
magnetic field can drive quasi-one-dimensional subsystems with a spin gap into
a gapless regime,
thus inducing long-range three-dimensional magnetic ordering due to
weak residual magnetic coupling between the subsystems.
Compounds with higher spin degrees of freedom, such as 
N-leg spin-1/2 ladders, are shown to have
cascades of ordering transitions.
At high magnetic fields,
zero-point fluctuations within the 
quasi-1D subsystems are suppressed, causing
quantum corrections to the ordering temperature to be reduced. 
\end{abstract}


Compounds with strongly anisotropic crystal structures typically 
exhibit low-dimensional behavior at high temperatures. At 
low temperatures, the specific nature of their quantum fluctuations
determines whether 
three-dimensional (3D)
ordering \cite{tennant} or other instabilities \cite{hase} occur, or 
whether there is
no transition at all \cite{imry}. In particular, weakly coupled
Heisenberg spin-1/2 chains are known to have a
3D magnetic ordering 
transition \cite{tennant} or a spin-Peierls (SP) 
instability at low temperatures \cite{hase},
if there is sufficiently strong coupling with low-lying 
phonon modes. On the other hand, compounds with an intrinsic spin gap, 
like e.g.\ 
weakly coupled integer-spin chains \cite{ma} or even-leg spin-1/2 
Heisenberg ladders in a spin-liquid state, 
retain their one-dimensionality down to
zero temperature\cite{azuma}. 
The pure RVB nature of their groundstate renders them
inert to weak residual magnetic couplings between the 
quasi-1D subsystems. Thus 3D magnetic ordering and
SP transitions are suppressed.

In this second class of materials, 
an applied magnetic field, $h$, can decrease the singlet-triplet
excitation gap of the quasi-1D subsystem, and eventually drive it
into a gapless regime if the field exceeds a critical strength, $h_{c1}$. 
A transition to a low-temperature ordered phase due to residual
magnetic couplings becomes again possible in this partially polarized
regime \cite{haas,giamarchi}. In this paper, we propose 
that 3D ordering as a result of the deconfinement
of pairs of bound spinons by an external 
magnetic field can actually be realized in a
wide variety of quasi-1D physical systems, including anisotropic spin chains,
ladders, and SP compounds. Furthermore, a unified phenomenology
for the magnetic phase diagram of these materials is presented,
starting from an analysis of weakly coupled antiferromagnetic
Heisenberg spin-1/2 chains (AFHC) with an easy-axis anisotropy.
Interestingly, we observe that for compounds
such as N-leg spin ladders with plateaus in their 
magnetization curves, $m(h)$,\cite{cabra}   
a cascade of N/2 ordering 
transitions for N even and (N+1)/2 transitions for N odd
occurs at high magnetic fields.  

We are especially interested in the application of this magnetic field
induced ordering transition to the material $Cu_2(C_2H_{12}N_2)_2Cl_4$
($CuHpCl$) \cite{chaboussant}.
Typically, the spin gap in most of the ladder compounds known to date
is too large to be
overcome by presently available magnetic fields.
However, 
this particular material has only a small spin gap of $\approx$ 10.5 Kelvin
which makes the interesting gapless regime experimentally accessible.
It has recently been pointed out that $CuHpCl$ may better be modeled
by an ensemble of weakly coupled dimers than as an antiferromagnetic
2-leg ladder \cite{broholm}. 
Whatever the precise magnetic structure may turn out to be, 
a magnetic field
induced ordering transition can occur in  all  
quasi-1D
spin systems with a singlet-triplet excitation gap, including Ising-like
chains, spin-Peierls chains, and ensembles of spin dimers.
Other possible candidate 
materials with (relatively small) spin gaps
include $KCuCl_3$ \cite{cavadini}, $CuGeO_3$ \cite{hase}, 
$\alpha'-NaV_2O_5$\cite{augier}, 
and the homologous series of cuprates $Sr_{n-1}Cu_{n+1}O_{2n}$ \cite{azuma}.

\begin{figure}
\centerline{\psfig{figure=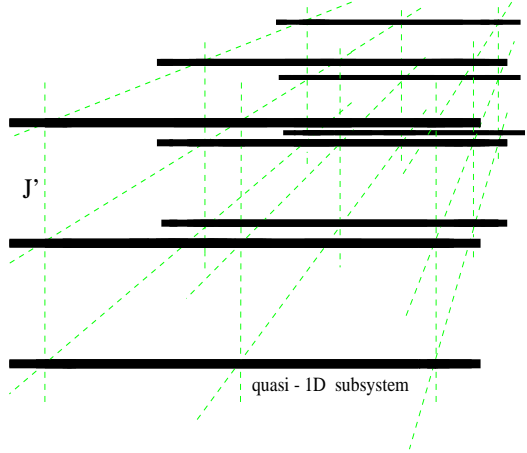,width=7cm,height=6.0cm,angle=0}}
\vspace{0.5cm}
\caption{
Schematic illustration of a strongly anisotropic 3D crystal, containing
quasi-1D subsystems, e.g.\ antiferromagnetic
spin ladders. 
The coupling $J'$ between the 1D
elements is small compared to the exchange constant $J$ within the 
subsystems. 
Therefore the compound has quasi-1D properties
at high temperatures ($T \gg J'$), while at low temperatures 
a 1D to 3D ordering transition can occur if the 1D subsystems 
are (driven) gapless. 
}
\end{figure}

Let us first consider a crystal of weakly coupled
anisotropic antiferromagnetic spin-1/2 Heisenberg chains, described by the 
Hamiltonian
\bea
H^{1D} = \sum_i \left[ J (S^x_iS^x_{i+1} + S^y_iS^y_{i+1} 
+ \Delta S^z_iS^z_{i+1}) - h S^z_i \right] ,
\eea
where $J > 0$ is the  antiferromagnetic
exchange constant within the chains, $\Delta$ is an easy-axis anisotropy,
and $h$ is an applied external magnetic field. 
The chains are weakly coupled by $H' = J'\sum_{<i,j>} {\bf S}_i
\cdot {\bf S}_j $ with $0 < J' \ll J$.
At zero magnetic field, the excitation spectrum  of $H^{1D}$
is gapless in the XY regime
($\Delta <  1$)
and massive in the Ising regime ($\Delta >  1$), with a Kosterlitz-Thouless
transition at the Heisenberg point ($\Delta =1$)\cite{luther}. 
In the Ising regime, a 
finite critical magnetic field, $h_g$, has to be overcome to 
completely soften the 
lowest triplet mode at wavevector $\pi$, 
and to drive the system gapless.

From a numerical solution of the Bethe Ansatz integral equations of Eq. 1 
\cite{bogoliubov}, 
we have obtained the magnetic field dependent spinon velocity, $u(h)$, the 
Luttinger exponent, $K(h)$, the magnetization, $m(h)$,
and the radius of compactification, $R(h)$, for the effective low-energy
c=1 conformal field theory. Magnetization curves, $m(h)$, for an
XY-like and an Ising-like anisotropic Heisenberg chain are shown in
Fig. 2. In the dispersion curves of the equivalent
spinless-fermion model (insets of Fig. 2),
the external magnetic field corresponds to a 
chemical potential. 
Below $h_{c1} = - J(1 + \Delta)$ the cosine band, $\omega (k)$, of 
the spinless fermions is empty, and $u(h_{c1}) = \left( \partial \omega
/\partial k \right)_{h_{c1}} 
 = 0$. 
At the special field $h=0$, the quasi-long-range ordered spin
density wave within the chains becomes commensurate
with the lattice, and Umklapp processes open up a spin gap
for $\Delta > 1$, leading to a plateau in the magnetization curve
(Fig. 2 (b)) for $-h_g<h<h_g$, with
\bea
h_g= 2 \pi J\frac{\sinh\gamma}{\gamma}
\sum_{n=0}^{\infty} \left[\cosh\left(\frac{
\left(2n+1\right)\pi^2}{2\gamma}\right) \right]^{-1},
\eea
where $\Delta=\cosh(\gamma)$ \cite{cloizeaux}.
Beyond $h_{c2} =  J(1 + \Delta)$, the chains are fully 
polarized in the direction of the applied magnetic field.
The
particle-hole symmetry about $h=0$ is reflected by the shape
of $m(h)$ and the 3D ordering temperature, $T_c(h)$ (Fig. 3),
discussed below. 

\begin{figure}
\centerline{\psfig{figure=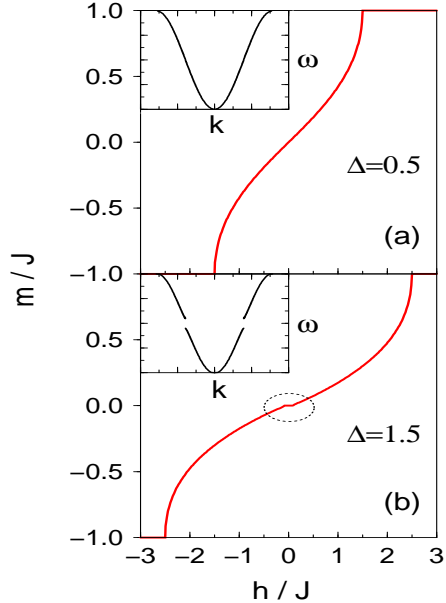,width=6cm,height=8.0cm,angle=270}}
\vspace{0.2cm}
\caption{
Magnetization curve, $m(h)$, of an antiferromagnetic spin-1/2 XXZ 
chain. 
(a) XY regime with Ising anisotropy: $\Delta = 0.5$,
and (b) Ising regime with $\Delta = 1.5$. The
insets show the dispersion law, $\omega (k)$, 
for the corresponding continuum 
models.
Here the magnetic field acts as a chemical potential, rising from
the bottom of the band at $h_{c1}$ to the top at $h_{c2}$. In the Ising
regime, a spin gap opens around h=0, leading to a
plateau in $m(h)$.}
\end{figure}

The low-temperature behavior of the susceptibility in the
gapless regime is determined by the dominant low-frequency
spinon modes at momentum $q_{z} = \pi $:
\begin{eqnarray}
\chi^{1D}_{+-}(q_{z},\omega =0; T)= 
F(\Delta)\left[\frac{\sin(\frac{\pi}{4K})}{u} 
\left(\frac{2\pi T}{u}\right)^{\frac{1}{2K}-2}\right. 
\nonumber \\
 \times \left.B^2\left( \frac{1}{8K},
1 - \frac{1}{4K} \right)-\frac{\pi}{u(1-1/4K)}\right],
\end{eqnarray}
where $B(x,y)$ is Euler's beta function,
and $F(\Delta)$ is a prefactor which strongly depends on the Ising anisotropy 
\cite{schulzlukyanov}.
Here, it has been assumed
that the chains are parallel to the $z$-axis of
the crystal. 
Furthermore, we have neglected higher-order logarithmic corrections which arise
in a more rigorous treatment of the backscattering processes.

\begin{figure}
\centerline{\psfig{figure=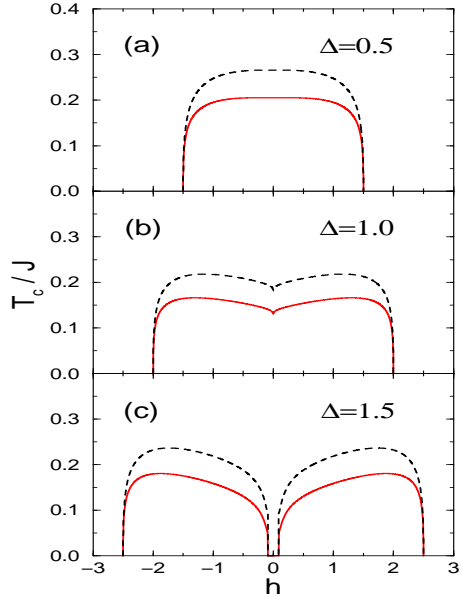,width=6cm,height=8.0cm,angle=270}}
\vspace{0.2cm}
\caption{
3D ordering temperature (solid line)
as a function of the applied magnetic field 
in a cubic crystal of weakly coupled antiferromagnetic spin-1/2
chains. (a) XY regime with $\Delta = 0.5$, (b) Heisenberg point
($\Delta = 1.0$),  and Ising regime with $\Delta = 1.5$.
The dashed lines indicate the onset of the fluctuation region 
below which the 3D magnetic 
correlation length becomes comparable to the inter-chain spacing.
For this plot, we have chosen $J'/J = 1/16$.
}
\end{figure}

Within the mean field approximation of Ref. \cite{schulz},
the low-temperature divergence in $\chi^{1D}( q_z,\omega =0; T)$ drives 
a 3D ordering transition due to the weak inter-chain couplings, $J'$: 
\begin{eqnarray}
\chi^{3D}({\bf q},\omega =0; T) = \frac{\chi^{1D}( q_{z},\omega =0; T)}
{1+J' f({\bf q}) \chi^{1D}( q_{z},\omega =0; T) },
\end{eqnarray}
where $f({\bf q})$ is the crystal form factor
which we here set to $f({\bf q})=-1$ for simplicity (simple cubic lattice).
The transition temperature, $T_c$, is determined from the locus of
divergence
of $\chi^{3D}({\bf q},\omega =0; T)$.
In Fig. 3, the magnetic field dependence of $T_c$ is shown in the critical
regime ($\Delta = 0.5$), at the marginal point ($\Delta = 1$), and in the
massive region ($\Delta = 1.5$). As the easy-axis anisotropy -
equivalent to
a nearest-neighbor repulsion
in the spinless fermion picture - is increased,
Umklapp scattering processes lead to a suppression of
$T_c(h)$ around $h = 0$. In the Ising regime these processes become relevant,
and a spin gap opens up, causing $T_c$ to vanish for $|h| < h_{g}
$.

To scrutinize the quality of this approach, we have calculated the 
Gaussian fluctuations which decrease the value of the transition temperature,
$T_c(h)$, obtained from the mean field calculation \cite{ginzburg}.
This suppression is found to be rather small (${\mathcal{O}}( 1 \% )$ )
because the corresponding Ginzburg-Landau parameter is proportional 
to the ratio
$J'/J \ll 1$. We thus conclude that the mean field treatment gives a 
quantitatively adequate description of the ordering transition. 
Furthermore, the onset of the fluctuation region below which 
the 3D magnetic correlation length becomes comparable to the inter-chain 
lattice spacing is indicated by
the dashed line in Fig. 3. This precursor region is quite large  
(${\mathcal{O}}( \approx 20 \% )$ ), indicating a sizable staggered 
magnetization, $m_s$, in the ordered regime. Clearly, zero-point fluctuations
are exponentially suppressed by $J'$.

Let us now examine magnetic field induced 3D ordering transitions
in quasi-1D systems with additional internal degrees of freedom,
such as weakly coupled Heisenberg spin-S chains with S $> 1/2$, N-leg S=1/2
ladders with N$\geq$2, and spin-Peierls
compounds. 

The low-energy Hamiltonian for
a SP chain in a magnetic field is given by
\bea 
H^{SP} =  \sum_i \left[ J\left( 1 + \delta (-1)^i \right) 
{\bf S}_i \cdot {\bf S}_{i+1} - h S^z_i \right] ,
\eea
where $\delta$ is the effective lattice distortion due to 
the coupling between
spinon and phonon degrees of freedom.
There are three well-known regimes in this system:
(i) for $|h|<h_{c1}$, it is in a 
spin-liquid phase with a
spin gap $\Delta_{SG} \simeq 
\delta^{2/3} |\log{\delta}|^{-1/2}$\cite{klumper} for sufficiently
small $\delta$'s, and  $h_{c1}= {\mathcal{O}}(\Delta_{SG}) $; 
(ii) for $h_{c1}<|h|<h_{c2}$ the system is a gapless spin-density wave with
a field-dependent modulation; 
(iii) for $|h|>h_{c2}$ it is fully polarized
in the direction of the applied magnetic field.

In the gapless, partially polarized, region (ii), 
we can map the low-energy spectrum
of this system 
onto an effective XY-like Heisenberg chain (Eq. 1) with the 
parameters: $J_{eff} = J (1 -\delta )/2 $, $\Delta_{eff} = 1/2 $, and 
$h_{eff} = h - J (5 + 3 \delta )/4$. This mapping is valid for 
$\delta \alt 1$. The lower and upper
critical fields of the gapless regime are
$ h_{c1} = J (1 + 3 \delta )/2 $  
and $ h_{c2} = 2 J $\cite{footnote2}. Considering weak antiferromagnetic 3D
couplings between the SP chains, 
$H' = J'\sum_{<i,j>} {\bf S}_i
\cdot {\bf S}_j $, the qualitative magnetic field dependence of 
the 3D ordering temperature, $T_c(h)$, is thus found to be 
the same as in
Fig. 3(a). 

As seen in the above example, 
weakly coupled anisotropic Heisenberg
chains in a field can serve as an effective theory for other 
quasi-1D quantum magnets with richer internal structure. 
Let us now extend this idea to weakly coupled N-leg spin-1/2 Heisenberg
ladders with the Hamiltonian
\bea 
H^N = 
J_{\parallel} \sum_{\leftrightarrow} {\bf S}_{i,\tau} \cdot {\bf S}_{j,\tau}
    + J_{\perp}\sum_{\updownarrow} {\bf S}_{i,\tau} \cdot {\bf S}_{i,\tau '},
- h \sum_{i,\tau } 
S^z_{i,\tau },
\eea
where i and j enumerate the rungs, $\rm \tau$, $\rm \tau '$
label the legs, and the sum marked by
$\leftrightarrow$ ($\updownarrow$) runs over
nearest neighbors along legs (rungs). 
Ladders with even and odd number of legs are known to have
quite distinct features at zero field ($h = 0$):
at low energies, 
odd-leg systems can be mapped onto effective gapless
antiferromagnetic Heisenberg chains with
longer-range interactions \cite{frischmuth}.
In even-leg systems,
relevant inter-band scattering processes open up a spin gap for any finite
positive
inter-chain coupling, $J_{\perp} > 0$ \cite{strong}
This is the reason why approximations based on
strong coupling anisotropies, i.e.\ expansions in $J_{||}/J_{\perp}$ 
\cite{cabra,milatandon}, give a
correct qualitative picture, extending even beyond the isotropic
($J_{||} = J_{\perp}$) regime\cite{cabra,strong}.
Furthermore, many physical
ladder materials show coupling anisotropies within the ladder
complex, e.g.\ a recent structural analysis of the 
candidate vanadate ladder material
$NaVa_2O_5$ suggests a strong rung-coupling anisotropy of
$J_{\parallel}/J_{\perp}\approx 13/75$ 
\cite{horsch}, and for the proposed 2-leg cuprate ladder,
 $CuHpCl$, the 
anisotropy ratio was estimated to be $J_{\parallel}/J_{\perp} \approx
1/5$ \cite{hayward}.  

It has recently been shown that N-leg Heisenberg spin-1/2 ladders can
have up to N/2 plateaus in their magnetization curve, $m(h)$, if N is even,
and (N+1)/2 plateaus if N is odd \cite{cabra}. 
These plateaus are related to gaps 
between the spin multiplets in the excitation spectra.
The partially polarized regions between the plateaus are gapless, and 
can be described by 
effective massless c=1 conformal field theories, 
corresponding to a spin-1/2 AFHC Hamiltonian with
parameters
$(J_{eff}, \Delta_{eff}, h_{eff} )$. 
In table 1, the numerical values of these parameter sets, 
obtained from a strong rung expansion ($J_{\parallel}/J_{\perp}
\ll 1$) are listed for N = 2, ..., 8 legs. 

\begin{table}
\caption{Parameters of the effective low-energy model 
for the gapless regions of N-leg spin-1/2 ladders in a magnetic field, $h$.
The effective magnetic field is given by 
$h_{eff}=h-h_c(0) J_{\perp}-c_h J_{\parallel}.$ }
\begin{tabular}{|l|l|l|l|l|}
$N$  & $J_{eff}/J_{\parallel}$ & $\Delta_{eff}$ & $h_c(0)$ & $c_h$\\ \hline
2& 1    & 0.5   & 1     & 0.5\\ \hline
3& (1,1)        & (1,0.5)       & (0,1.5)     & (0,0.5)  \\ \hline
4& (1.075,1)& (0.3489,0.375)& (0.6589,1.7071)&
(0.375,0.625)\\ \hline
5& (1.0169,1.0961,1)& (1,0.3789,0.3)& (0,1.1189,1.809)&
(0,0.2958,0.7)\\ \hline
6& (1.1114,1.1348,1)& (0.3163,0.3011,0.25)&
(0.4916,1.386,1.866)&
(0.3515,0.3985,0.75)\\ \hline
7& (1.0344,1.1415,& (1,0.3407,&
(0,0.8848,& (0,0.2166,\\
& 1.1663,1)& 0.2381,0.2143)&
1.5504,1.901)& 0.4891,0.7857)\\ \hline
8& (1.1364,1.1882& (0.302,0.2743,&
(0.3926,1.1506,& (0.3432,0.2888,\\
& 1.1917,1)& 0.1962,0.1875)&
1.6577,1.9239)& 0.5555,0.8125)\\ 
\end{tabular}
\end{table}

Applying the mean field approach described above to 
a crystal of weakly
coupled N-leg ladders, we now observe a cascade of N/2 ((N+1)/2)
3D ordering transitions for quasi-1D ladder subsystems with an
even (odd) number of legs, shown in Fig. 4.
In the case of weakly coupled even-leg ladders, the first transition
is driven by the formation of a quasi-long-range
ordered SDW of triplets
along the ladder direction, commensurate with the magnetic field, $h > h_{c1}$.
The following transition (for $ N > 2$) is driven by a SDW of
quintuplets, etc. . Depending on the ratio $J_{\parallel}/J_{\perp}$,
these phases of different multiplet polarization may overlap,
and mixed phases can occur (see  e.g.\
Fig. 4 (d) around $h = 1.5 J_{\perp}$ and $1.8 J_{\perp}$).
The resulting 3D ordering temperature does not vanish completely in this 
case, but has minima at particular magnetic fields where the number  
of the lower multiplet excitations equals the number of the 
next-higher multiplet excitations.
By analogy, a cascade in $T_c(h)$ should occur in compounds 
containing weakly coupled integer-spin  Heisenberg spin chains.

\begin{figure}
\centerline{\psfig{figure=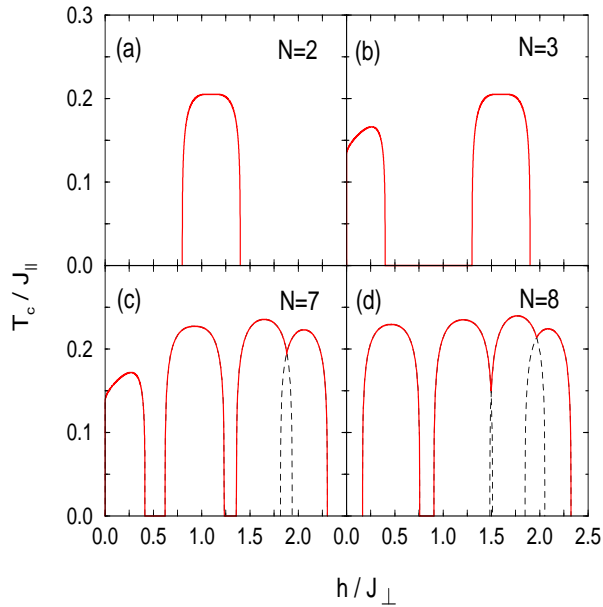,width=8.0cm,height=8.0cm,angle=270}}
\vspace{0.5cm}
\caption{
3D ordering transition temperatures of N-leg spin-1/2 Heisenberg 
ladders as a function of an external magnetic field. Cascades of
transitions are observed for $N > 2$, driven by 1D SDW's
of spin-S multiplets on the ladders.
For this plot, we have chosen an anisotropy ratio $J_{\parallel}/J_{\perp}
= 1/5$ and residual inter-ladder couplings $J'/J_{\parallel} = 1/16$.
}
\end{figure}

Odd-leg spin-1/2 ladders  - as well as half-odd-integer-spin
Heisenberg chains -
also have a sequence of ordering transitions, with the only difference that
the onset of the first transition occurs already at $h = 0$ (Figs. 
4 (b) and (c)). The dip features in $T_c(h)$, due to strong Umklapp 
scattering  ($\Delta_{eff} \alt 1$) become less pronounced at larger
magnetic fields, i.e.\ $\Delta_{eff} \rightarrow 0$. 
This is expected because the effective spin degrees of freedom of the
higher multiplets are more ``classical"  as $S \rightarrow \infty$,
and quantum fluctuations are suppressed at high fields.
Finally, as the width of the ladder subsystems is increased, 
the gaps between the various multiplets 
disappear, and a quasi-continuous  $T_c(h)$-curve emerges  - as 
it is expected for the limiting case ($N = \infty$) of weakly coupled 
2D Heisenberg planes.

In summary, we have examined 3D magnetic ordering
transitions in quasi-1D materials.
In the presence of a spin gap, an external
 magnetic field can drive the 1D subsystems
gapless, thus inducing long-range 3D magnetic ordering due to
residual magnetic coupling between the subsystems.
For compounds with higher spin degrees of freedom, such as 
N-leg spin-1/2 ladders, cascades of ordering transitions are
predicted to occur, reflecting the mesoscopic
character of these materials. At high magnetic fields,
quantum fluctuations in the 
quasi-1D subsystems are suppressed, causing
the ``dip feature" in $T_c(h)$ -  due to Umklapp processes - 
to disappear for the transitions at higher fields.

We wish to thank 
G. Bickers, C. Broholm, E. Dagotto, A. Honecker, D. Scalapino,
and A. Tsvelik
for useful discussions,
and acknowledge
the Zumberge Foundation for financial support. 
S.W. is partially supported by DAAD
grant No. HSP III,D/98/11174.


\begin{references}
\bibitem{footnote2} In the case of an additional next nearest neighbor 
interaction, $J_2$, the parameters are $J_{eff}=J(1-\delta)-J_2$, 
$\Delta_{eff}=(J(1-\delta)+J_2)/2(J(1-\delta)-J_2)$, 
$h_{eff}=h-J(5+3\delta)/4-J_2/2$.
\end{references}
\end{document}